\documentclass[11pt]{article}
\usepackage{amsmath,amssymb,fullpage}
\input{epsf}

\newcommand{\ignore}[1]{}
\newtheorem{theorem}{Theorem}
\newtheorem{theoremone}{Theorem}
\newtheorem{lemma}{Lemma}
\newtheorem{claim}{Claim}
\newtheorem{proposition}{Proposition}

\newtheorem{corollary}{Corollary}

\newenvironment{proof}{{\bf Proof.}}{\hfill\rule{2mm}{2mm}}

\newenvironment{proofof}[1]
      {\medskip\noindent{\bf #1 :}\hspace{1em}}
      {\hfill\rule{2mm}{2mm}}

\newtheorem{remarka}{Remark}

\newtheorem{prelem}{{\bf Theorem}}

\newtheorem{preconj}{{\bf Conjecture}}

\def\Min {{\rm \bf Min}}

\def\sd {{\rm sd}}
\def\vc {{\rm vc}}
\def\bo {{\bf 0}}
\def\bu {{\bf u}}
\def\bv {{\bf v}}
\def\bw {{\bf w}}
\def\by {{\bf y}}
\def\eps {\epsilon}
\def \R {\mathbb{R}}
\title{Integrality gaps of semidefinite programs for Vertex Cover and
relations to $\ell_1$ embeddability of negative type metrics}
\author{
Hamed Hatami
\and
Avner Magen
\and
Evangelos Markakis
\and
\\ \small{Department of Computer Science}
\\ \small{University of Toronto}
}

\date{}
\begin{document}

\maketitle
\begin{abstract}
We study various SDP formulations for {\sc Vertex Cover} by adding
different constraints to the standard formulation. We show that
{\sc Vertex Cover} cannot be approximated better than $2-o(1)$
even when we add the so called pentagonal inequality constraints to the
standard SDP formulation, en route answering an open
question of Karakostas~\cite{Karakostas}. We further show the
surprising fact that by strengthening the SDP with the (intractable) 
requirement that the metric interpretation of the solution is an $\ell_1$ 
metric, we get an exact relaxation (integrality gap is 1), and on the other 
hand if the solution is arbitrarily close to being $\ell_1$ embeddable, the 
integrality gap may be as big as $2-o(1)$. 
Finally, inspired by the above findings, we use ideas from the integrality gap
construction of Charikar \cite{Char02} to provide a family of simple 
examples for negative type metrics that cannot be embedded into $\ell_1$ 
with distortion better than $8/7-\eps$. To this end we prove a new 
isoperimetric inequality for the hypercube.
\end{abstract}
\thispagestyle{empty} 

\pagebreak

\setcounter{page}{1}

\section{Introduction}

A {\sf vertex cover} in a graph $G=(V,E)$ is a set $S \subseteq V$
such that every edge $e \in E$ intersects $S$ in at least one
endpoint. Denote by $\vc(G)$ the size of the minimum vertex cover
of $G$. It is well-known that the minimum vertex cover problem has a
$2$-approximation algorithm, and it is widely believed that for
every constant $\eps>0$, there is no
($2-\eps$)-approximation algorithm for this problem. Currently the
best known hardness result for this problem shows that
$1.36$-approximation is NP-hard \cite{DS02}. If we were to assume
the Unique Games Conjecture~\cite{Khot}, the problem would be
essentially settled as $2-\Omega(1)$ would then be NP-hard
\cite{KhotRegev}. 

In a seminal paper, Goemans and Williamson \cite{GW95}
introduced semidefinite programming as a tool for obtaining
approximation algorithms. Since then semidefinite programming has
been applied to various approximation problems and has become an
important technique, and indeed the best known approximation
algorithms for many problems are obtained by solving an SDP
relaxation of them.

The best known algorithms for {\sc Vertex Cover} compete in ``how big
is the little o'' in the $2-o(1)$ factor. The best two are in fact based on
SDP relaxations: Halperin \cite{H02} gives a  
$(2-\log \log \Delta / \log \Delta)$-approximation where $\Delta$ is the 
maximal degree of the graph while Karakostas obtains
a $(2-\Omega(1/\sqrt{\log n}))$-approximation \cite{Karakostas}.

The standard way to formulate the {\sc Vertex Cover} problem as a quadratic 
integer program is the following: 
\begin{equation*}
\begin{array}{clcl}
\Min & \sum_{i \in V} (1+x_0 x_i)/2 &\qquad &\\
{\rm s.t.} &  (x_i-x_0)(x_j-x_0) = 0  & &\forall \ ij \in E \\
& x_i \in \{-1,1\}  & & \forall \ i \in \{0\} \cup V,
\end{array}
\end{equation*}
where the set of the vertices $i$ for which $x_i=x_0$ correspond
to the vertex cover. By relaxing this integer program to a
semidefinite program, the scalar variable $x_i$ now becomes a vector $\bv_i$ and we get:
\begin{equation}
\label{SDP-VC1}
\begin{array}{clcl}
\Min & \sum_{i \in V} (1+\bv_0\bv_i)/2 &\qquad &\\
{\rm s.t.} &  (\bv_i-\bv_0)\cdot(\bv_j-\bv_0) = 0  & &\forall \ ij \in E \\
&\| \bv_i \| = 1  & & \forall \ i \in \{0\} \cup V.
\end{array}
\end{equation}
Kleinberg and Goemans~\cite{KG95} proved that SDP (\ref{SDP-VC1})
has integrality gap of $2-o(1)$. Specifically, given $\epsilon > 0$, they construct a graph $G_\eps$ for which $\vc(G_\eps)$ is at least $(2-\eps)$ times 
larger than the optimal solution to the SDP. They also suggested the following
strengthening of SDP (\ref{SDP-VC1}) and left its integrality gap
as an open question:
\begin{equation}
\label{SDP-VC2}
\begin{array}{clcl}
\Min & \sum_{i \in V} (1+\bv_0\bv_i)/2 &\qquad &\\
{\rm s.t.} &  (\bv_i-\bv_0)\cdot (\bv_j-\bv_0) = 0  & &\forall \ ij \in E \\
&  (\bv_i-\bv_k)\cdot (\bv_j-\bv_k) \ge 0  & &\forall \ i,j,k \in \{0\} \cup V \\
&\| \bv_i \| = 1  & & \forall \ i \in \{0\} \cup V.
\end{array}
\end{equation}
Charikar~\cite{Char02} answered this question by showing that 
the same graph $G_\eps$ but a different vector solution satisfies SDP 
(\ref{SDP-VC2})\footnote{To be more precise, Charikar's proof was for a slightly weaker formulation than (\ref{SDP-VC2}) 
but it is not hard to see that the same construction
works for SDP (\ref{SDP-VC2}) as well.} 
and gives rise to an integrality gap of $2-o(1)$ as before.
The following is an equivalent formulation to SDP (\ref{SDP-VC2}):
\begin{equation}
\label{SDP-VC3}
\begin{array}{clcl}
\Min & \sum_{i \in V} 1-\|\bv_0-\bv_i\|^2/4 &\qquad &\\
{\rm s.t.} &  \|\bv_i-\bv_0\|^2+\|\bv_j-\bv_0\|^2 = \|\bv_i-\bv_j\|^2  & &\forall \ ij \in E \\
 &  \|\bv_i-\bv_k\|^2+\|\bv_j-\bv_k\|^2 \ge \|\bv_i-\bv_j\|^2  & &\forall \ i,j,k \in \{0\} \cup V \\
&\| \bv_i \| = 1  & & \forall \ i \in \{0\} \cup V \\
\end{array}
\end{equation}

\paragraph{Viewing SDPs as relaxations over $\ell_1$} 
The above reformulation reveals a connection to 
metric spaces. The second constraint in SDP (\ref{SDP-VC3}) says
that $\|\cdot\|^2$ induces a metric on $\{\bv_i:i \in \{0\} \cup V\}$,
while the first says that $\bv_0$
is on the shortest path between the images of every two
neighbours. This suggests a more careful study of the problem from
the metric viewpoint which is the purpose of this article. Such
connections are also important in the context of the {\sc Sparsest Cut}
problem, where the natural SDP relaxation was analyzed in the
breakthrough work of Arora, Rao and Vazirani \cite{ARV} and it was
shown that its integrality gap is at most $O(\sqrt {\log n})$. This later
gave rise to some significant progress in the theory of metric spaces
\cite{CGR,ALN}.

For a metric space $(X,d)$, let $c_1(X,d)$ denote the minimum
distortion required to embed $(X,d)$ into $\ell_1$ (see
\cite{Mat02} for the related definitions). So $c_1(X,d)=1$ if and
only if $(X,d)$ can be embedded isometrically into $\ell_1$.
Consider a vertex cover $S$ and its corresponding solution to SDP
(\ref{SDP-VC2}), i.e., $\bv_i=1$ for every $i \in S \cup \{0\}$
and $\bv_i=-1$ for every $i \not\in S$. The metric defined by
$\|\cdot\|^2$ on this solution (i.e.,
$d(i,j)=\|\bv_i-\bv_j\|^2$) is isometrically embeddable
into $\ell_1$. Thus we can strengthen SDP (\ref{SDP-VC2}) by
allowing any arbitrary list of valid inequalities in $\ell_1$ to be added. 
For example the triangle inequality, is one 
type of such constraints. The next natural inequality of this
sort is the {\sf pentagonal inequality}: A metric space $(X,d)$ is
said to satisfy the pentagonal inequality if for $S,T \subset X$
of sizes 2 and 3 respectively it holds that $\sum_{i\in S,j\in
T}d(i,j) \ge \sum_{i,j\in S}d(i,j) + \sum_{i,j\in T}d(i,j)$. Note that this
inequality does no longer apply to every metric, but it does to  ones
that are $\ell_1$ embeddable. This leads to the following natural 
strengthening 
\ignore{
\footnote{It is easy to see that if two points from $S$ and $T$
are the same then the pentagonal inequality reduces to triangle
inequality on the remaining 3 points}}
of SDP (\ref{SDP-VC3}):
\begin{equation}
\label{SDP-VC4}
\begin{array}{clcl}
\Min & \sum_{i \in V} 1-\|\bv_0-\bv_i\|^2/4 &\qquad &\\
{\rm s.t.} &  \|\bv_i-\bv_0\|^2+\|\bv_j-\bv_0\|^2 = \|\bv_i-\bv_j\|^2  & &\forall \ ij \in E \\
&
\begin{array}{cl}
\sum_{i \in S,j \in T} \|\bv_i-\bv_j\|^2 \ge & \sum_{i,j \in S}\|\bv_i-\bv_j\|^2+\\
&\sum_{i,j \in T}\|\bv_i-\bv_j\|^2
\end{array}
& &
\begin{array}{l}
\forall \ S,T \subseteq \{0\} \cup V, \\ |S|=2, |T|=3
\end{array}
\\
&\| \bv_i \| = 1  & & \forall \ i \in \{0\} \cup V \\
\end{array}
\end{equation}
%

In Theorem~\ref{Thm:Pentagonal}, we prove that SDP
(\ref{SDP-VC4}) has an integrality gap of $2-\eps$, for every
$\eps>0$. It is interesting to note that for the classical problem of {\sc
Sparsest Cut}, it is not known how to show a nonconstant
integrality gap against pentagonal (or any other $k$-gonal)
inequalities, although recently a nonconstant integrality gap was
shown by Khot and Vishnoi and later by Devanur {\em et al.}
\cite{KhotVishnoi, DKSV06} in the presence of the triangle
inequalities\footnote{As Khot and Vishnoi note, and leave as an
open problem, it is possible that their example satisfies some or all
$k$-gonal inequalities.}.

One can actually impose any $\ell_1$-constraint not only for the metric defined by $\{\bv_i : i \in V \cup \{0\}\}$, but also for the one that comes from $\{\bv_i : i \in V \cup \{0\}\} \cup \{-\bv_i : i \in V\cup \{0\}\} $. 
This fact is used in \cite{Karakostas} where the triangle inequality 
constraints on this extended set are added, achieving an integraility gap 
of at most $2 -\Omega(\frac{1}{\sqrt{\log n}})$. It is also asked whether the 
integrality gap of this strengthening breaks the ``$2-o(1)$
barrier''. In Section~\ref{Sec:strongerSDP} we answer this question 
in the negative.

\paragraph{Integrality gap with respect to $\ell_1$ embeddability}
At the extreme, strengthening the SDP with $\ell_1$-valid constraints, 
would imply the condition that the
metric defined by $\|\cdot\|$ on $\{\bv_i:i\in \{0\} \cup V\}$, namely 
$d(i,j)=\|\bv_i-\bv_j\|^2$ is $\ell_1$ embeddable. Doing so leads to 
the following intractable program (which we refer to as SDP for convenience):

\begin{equation}
\label{SDP-l1}
\begin{array}{clcl}
\Min & \sum_{i \in V} 1-\|\bv_0-\bv_i\|^2/4 &\qquad &\\
{\rm s.t.} &  \|\bv_i-\bv_0\|^2+\|\bv_j-\bv_0\|^2 = \|\bv_i-\bv_j\|^2  & &\forall \ ij \in E \\
&\| \bv_i \| = 1  & & \forall \ i \in \{0\} \cup V \\
&c_1(\{\bv_i:i \in \{0\} \cup V\},\|\cdot\|^2) = 1 &&
\end{array}
\end{equation}

In \cite{ACMM05}, it is shown that an SDP formulation of {\sc
Minimum Multicut}, even with the constraint that the $\|\cdot\|^2$
distance over the variables is isometrically embeddable into
$\ell_1$, still has a large integrality gap. For the {\sc Max Cut}
problem, which is more intimately related to our problem, it is
easy to see that the $\ell_1$ embeddability condition does not
prevent the integrality gap of $8/9$; it is therefore tempting to
believe that there is a large integrality gap for SDP
(\ref{SDP-l1}) as well. 
Surprisingly, SDP (\ref{SDP-l1}) has no
gap at all; in other words, as we show in Theorem~\ref{Thm:SDP-l1},
the answer to SDP (\ref{SDP-l1}) is exactly the size of the
minimum vertex cover. A consequence of this fact is that any
feasible solution to SDP (\ref{SDP-VC2}) that surpasses the minimum vertex 
cover induces an $\ell_2^2$ distance which is not isometrically embeddable 
into $\ell_1$. This includes the integrality gap constructions 
of Kleinberg and Goemans', and that of Charikar's 
for SDPs (\ref{SDP-VC2}) and (\ref{SDP-VC3}) respectively. 
The construction of Charikar is more interesting in the sense that the 
obtained $\ell_2^2$ distance is a metric (from now on we refer to it 
as a {\sf negative type metric}; see \cite{dela97} for background and 
nomenclature). 
In contrast to Theorem \ref{Thm:SDP-l1}, we show in Theorem~\ref{Thm:SDP-weakl1} that if we relax the last constraint in SDP (\ref{SDP-l1}) to 
$c_1(\{\bv_i:i \in \{0\} \cup V\},\|\cdot\|^2) \le 1+\delta$ for any constant 
$\delta > 0$, then the integrality gap may ``jump'' to $2-o(1)$.
Compare this with a problem such as {\sc Sparsest Cut} in which 
an addition of such a constraint immediately implies integrality gap at most 
$1+\delta$.

\paragraph{Negative type metrics that are not $\ell_1$ embeddable}
Inspired by the above results, we construct in
Theorem~\ref{Thm:distortion} a simple negative type metric space
$(X,\|\cdot\|^2)$ that does not embed well into $\ell_1$.
Specifically, we get $c_1(X) \ge \frac{8}{7}-\eps$ for every
$\eps>0$. In order to show this we prove a new isoperimetric
inequality for the hypercube $Q_n = \{-1,1\}^n$, which we
believe is of independent interest. This theorem
generalizes the standard one, and under certain conditions provides
better guarantee for edge expansion:
\begin{theorem}
\label{Thm:isoper}(Generalized Isoperimetric inequality) For every
set $S \subseteq Q_n$,
$$|E(S,S^c)| \ge |S|(n-\log_2 |S|)+p(S).$$
where $p(S)$ denotes the number of
vertices $\bu \in S$ such that $-\bu \in S$. 
\end{theorem}

Khot and Vishnoi~\cite{KhotVishnoi} constructed an example of an $n$-point 
negative type metric that for every $\delta>0$ requires distortion at least
$(\log \log n)^{1/6-\delta}$ to embed into $\ell_1$. Krauthgamer
and Rabani \cite{KrauthgamerRabani} showed that in fact Khot and Vishnoi's
example requires a distortion of at least $\Omega(\log \log n)$.
Later Devanur {\em et al.} \cite{DKSV06} showed an example which suffers
an $\Omega (\log \log n)$ distortion even on average when embedded
into $\ell_1$ (we note that our example is also ``bad'' on
average). Although the above examples require nonconstant
distortion to embed into $\ell_1$, we believe that
Theorem~\ref{Thm:distortion} is interesting for the following
reasons: (i) Khot and Vishnoi's example is quite complicated, and
there is no good explanation to the fact that triangle inequality holds (citing the authors ``this is where the magic happens''). 
Simple constructions such as the one we obtain may give a better 
understanding of the problem and lead to simpler constructions of negative type
metrics that behave poorly in the above sense
(ii) there are not many known examples of negative type metrics that require a
constant  $c>1$ distortion to embed into $\ell_1$, and finding such examples 
is challenging and desirable. In fact before Khot and Vishnoi's
result, the best known lower bounds (see
\cite{KhotVishnoi}) were due to Vempala, $10/9$ for a metric
obtained by a computer search, and Goemans, $1.024$ for a metric
based on the Leech Lattice (compare these to the $8/7-\epsilon$
bound of Theorem~\ref{Thm:distortion}). We mention that by \cite{ALN} every negative type metric embeds into $\ell_1$ with distortion $O(\sqrt{\log{n}}\log{\log{n}})$.

\section{Preliminaries and notation \label{sec:notation}}
A vertex cover of a graph $G$ is a set of vertices that touch all edges. 
An independent set in $G$ is a set $I \subseteq V$ such that
no edge $e\in E$ joins two vertices in $I$. We denote by $\alpha(G)$ 
the size of the maximum independent set of $G$. Vectors are always denoted 
in bold font (such as $\bv$,
$\bw$, etc.); $\|\bv\|$ stands for the Euclidean norm of $\bv$, $\bu\cdot
\bv$ for the inner product of $\bu$ and $\bv$, and $\bu \otimes
\bv$ for their tensor product. Specifically, if $\bv, \bu \in
\R^n$, $\bu \otimes \bv$ is the vector with coordinates indexed by
ordered pairs $(i,j)\in [n]^2$ that assumes value $\bu_i \bv_j$ on
coordinate $(i,j)$. Similarly, the tensor product of more than two
vectors is defined. It is easy to see that $(\bu \otimes \bv) .
(\bu' \otimes \bv') = (\bu\cdot \bu') (\bv\cdot \bv')$. For two
vectors $\bu \in \R^n$ and $\bv \in \R^m$, denote by $(\bu,\bv)
\in \R^{n+m}$ the vector whose projection to the first $n$
coordinates is $\bu$ and to the last $m$ coordinates is $\bv$.

Next, we give a few basic definitions and facts about finite metric spaces. A 
metric space $(X,d_X)$ embeds with distortion at most $D$ into $(Y,d_Y)$ if
there exists a mapping $\phi : X \mapsto Y$ so that for all $a,b \in X$ 
$\gamma\cdot d_X(a,b) \le d_Y(\phi(a),\phi(b)) \le \gamma D \cdot d_X(a,b)$, for some $\gamma > 0$. We say that 
$(X,d)$ is $\ell_1$ embeddable if it can be embedded with distortion 1 into 
$\R^m$ equipped with the $\ell_1$ norm. An $\ell_2^2$ distance on $X$
is a distance function for which there there are vectors 
$\bv_x \in \R^m$ for every $x\in X$ so that $d(x,y) = \|\bv_x - \bv_y\|^2$. If, 
in addition, $d$ satisfies triangle inequality, we say that $d$ is an $\ell_2^2$ metric or {\sf negative type metric}. It is well known \cite{dela97} that
every $\ell_1$ embeddable metric is also a negative type metric.


\section{$\ell_1$ and Integrality Gap of SDPs for Vertex Cover  -- 
an ``all or nothing'' phenomenon} It is well known that for the {\sc Sparsest Cut} problem,
there is a tight connection between $\ell_1$ embeddability and
integrality gap.  In fact the integrality gap is bounded above by the
least $\ell_1$ distortion of the SDP solution.  At the other extreme
stand problems like {\sc Max Cut} and {\sc Multi Cut}, where $\ell_1$
embeddability does not provide any strong evidence for small
integrality gap. In this section we show that {\sc Vertex Cover}
falls somewhere between these two classes of $\ell_1$-integrality gap
relationship, and it witnesses a sharp transition in integrality gap in the
following sense: while $\ell_1$ embeddability prevents {\em any}
integrality gap, allowing a small distortion, say $1.001$ does not
prevent integrality gap of $2-o(1)$!

\begin{theorem}
\label{Thm:SDP-l1} For a graph $G=(V,E)$, the answer to the SDP
formulated in SDP (\ref{SDP-l1}) is the size of the minimum vertex
cover of $G$.
\end{theorem}
\begin{proof}
Let $d$ be the metric solution of SDP (\ref{SDP-l1}). We know that $d$ is 
the result of an $\ell_2^2$ unit representation (i.e., it comes from 
square norms between unit vectors), and furthermore it is $\ell_1$ 
embeddable.
By a well known fact about $\ell_1$ embeddable metrics (see, eg, \cite{dela97})
we can assume that there exist $\lambda_t>0$
and $f_t:\{0\} \cup V \rightarrow \{-1,1\}$, $t=1,\ldots,m$, such that
\begin{equation}
\label{distances} \|\bv_i-\bv_j\|^2=\sum_{t=1}^{m} \lambda_t
|f_t(i)-f_t(j)|,
\end{equation}
for every $i,j \in \{0\} \cup V$. Without loss of generality, we can assume that $f_t(0) = 1$ for every $t$. For convenience, we switch to talk about 
{\sc Independent Set} and its relaxation, which is the same as SDP 
(\ref{SDP-l1}) except for the objective function that becomes 
$\rm Max \sum_{i \in V} \|\bv_0-\bv_i\|^2/4$. Obviously, the theorem follows 
from showing that this is an exact relaxation.

We argue that 
(i) $I_t=\{i \in V: f_t(i)=-1\}$ is a (nonempty) independent set for 
every $t$, and 
(ii) $\sum \lambda_t = 2$.
Assuming these two statements we get
\begin{eqnarray*}
\sum_{i \in V} \frac{\|\bv_i - \bv_0\|^2}{4} &=&\sum_{i \in V}
\frac{\sum_{t=1}^{m} \lambda_t |1-f_t(i)|}{4}= \sum_{t=1}^m \frac{\lambda_t
|I_t|}{2} \le \max_{t \in [m]} |I_t| \le \alpha(G),
\end{eqnarray*}
and so the relaxation is exact and we are done.

We now prove the two statements. The first is rather straightforward: For $i,j \in I_t$, (\ref{distances}) implies that $d(i,0)+d(0,j)>d(i,j)$. It follows that $ij$ cannot
be an edge else it would violate the first condition of the SDP. (We may 
assume that $I_t$ is nonempty since otherwise the $f_t(\cdot)$ terms have no contribution in 
(\ref{distances}).)
The second statement is more surprising and uses the fact that the solution 
is optimal. The falsity of such a statement for the problem of 
{\sc Max Cut} (say) explains the different behaviour of the latter problem 
with respect to integrality gaps of $\ell_1$ embeddable solutions. We now describe the proof.

Let $\bv_i'=(\sqrt{\lambda_1/2}f_1(i),\ldots,\sqrt{\lambda_m/2}f_m(i),0)$.
From (\ref{distances}) we conclude that 
$\|\bv_i'-\bv_j'\|^2 = \|\bv_i-\bv_j\|^2$, hence there exists a vector 
$\bw = (w_1,w_2,...,w_{m+1})
\in \R^{m+1}$ and an orthogonal transformation $T$, such
that
$$\bv_i=T\left(\bv_i'+\bw\right).$$
Since the constraints and the objective function of the SDP
are invariant under orthogonal transformations,
without loss of generality we may assume that
$$\bv_i=\bv_i'+\bw,$$
for $i \in V \cup \{0\}$. We know that
\begin{equation}
\label{normv_i}
1=\|\bv_i\|^2=\|\bv_i'+\bw\|^2= w_{m+1}^2+\sum_{t=1}^m
(\sqrt{\lambda_t/2}f_t(i)+w_t)^2.
\end{equation}
Since $\|\bv'_i\|^2=\|\bv'_0\|^2=\sum_{t+1}^m \lambda_t/2$, for every $i
\in V \cup \{0\}$, from (\ref{normv_i}) we get
$\bv'_0 \cdot \bw=\bv'_i\cdot \bw$. Summing this over all $i \in V$, we have
$$|V|(\bv'_0\cdot \bw)=\sum_{i\in V}\bv'_i\cdot \bw
=\sum_{t=1}^m (|V|-2|I_t|)\sqrt{\lambda_t/2} w_t,$$
or
$$\sum_{t=1}^m |V|\sqrt{\lambda_t/2}w_t=\sum_{t=1}^m (|V|-2|I_t|)\sqrt{\lambda_t/2} w_t ,$$
and therefore
\begin{equation}
\label{sumzero} \sum_{t=1}^m |I_t|\sqrt{\lambda_t/2} w_t=0.
\end{equation}
Now (\ref{normv_i}) and (\ref{sumzero}) imply that
\begin{equation}
\label{coreineq}
\max_{t \in [m]} |I_t| \ge \sum_{t=1}^m
(\sqrt{\lambda_t/2}f_t(0)+ w_t)^2 |I_t| = \sum_{t=1}^m
\left(\frac{\lambda_t |I_t|}{2} + w_t^2|I_t|\right) \ge \sum_{t=1}^m
\frac{\lambda_t |I_t|}{2}.
\end{equation}
As we have observed before 
$$ \sum_{t=1}^m\frac{\lambda_t |I_t|}{2} = \sum_{i \in V} \frac{\|\bv_i - \bv_0\|^2}{4} $$ which means (as clearly $\sum_{i \in V} \frac{\|\bv_i - \bv_0\|^2}{4} 
\ge \alpha(G)$) that the inequalities in (\ref{coreineq}) must be tight. Now, 
since $|I_t|>0$ we get that $\bw = \bo$ and from (\ref{normv_i}) we get the 
second statement, i.e., $\sum \lambda_t = 2$.
This concludes the proof.
\end{proof}

Now let us replace the last constraint in SDP (\ref{SDP-l1}),
$c_1(\{\bv_i:i \in \{0\} \cup V\},\|\cdot\|^2) = 1$, with a weaker
condition $c_1(\{\bv_i:i \in \{0\} \cup V\},\|\cdot\|^2) \le
1+\delta$, for arbitrary $\delta>0$.

\begin{equation*}
\begin{array}{clcl}
\Min & \sum_{i \in V} 1-\|\bv_0-\bv_i\|^2/4 &\qquad &\\
{\rm s.t.} &  \|\bv_i-\bv_0\|^2+\|\bv_j-\bv_0\|^2 = \|\bv_i-\bv_j\|^2  & &\forall \ ij \in E \\
&\| \bv_i \| = 1  & & \forall \ i \in \{0\} \cup V \\
&c_1(\{\bv_i:i \in \{0\} \cup V\},\|\cdot\|^2) \le 1+\delta &&
\end{array}
\end{equation*}

\begin{theorem}
\label{Thm:SDP-weakl1} For every $\eps>0$, there is a graph
$G$ for which $\frac{\vc(G)}{\sd(G)} \ge 2-\eps$, where
$\sd(G)$ is the solution to the above SDP 
\end{theorem}

For the proof we show that the negative type metric implied by
Charikar's solution (after adjusting the parameters appropriately) 
requires distortion of at most $1+\delta$.
We postpone the proof to the appendix.


\section{Integrality Gap against the stronger Semi Definite formulations}
In this section we discuss the integrality gap for stronger
semi-definite formulations of vertex cover. In particular we show
that Charikar's construction satisfies both SDPs
(\ref{SDP-Karakostas}) and (\ref{SDP-VC4}). We start by describing
this construction.

\subsection{Charikar's construction \label{Charikar}}
The graphs used in the construction are the so called Hamming
graphs. These are graphs with vertices $\{-1,1\}^n$ and two vertices
are
 adjacent if their Hamming distance is exactly an even integer
$d=\gamma n$. A result of Frankl and R\"{o}dl \cite{FR87} shows
 that
$\vc(G) \ge 2^n-(2-\delta)^n$, where $\delta>0$ is a constant
depending only on $\gamma$. Kleinberg and Goemans \cite{KG95}
 showed
that by choosing proper $n$ and $\gamma$, this graph gives
 an
integrality gap of $2-\eps$ for SDP (\ref{SDP-VC1}). Charikar
\cite{Char02} showed that in fact $G$ implies the same result for
the SDP formulation in (\ref{SDP-VC2}) too. To this end he
introduced the following solution to SDP (\ref{SDP-VC2}):

For every $\bu_i \in \{-1,1\}^n$, define
$\bu_i'=\bu_i/\sqrt{n}$,
so that $\bu_i'\cdot \bu_i'=1$. Let $\lambda=1-2\gamma$, $q(x) =
x^{2t} + 2t\lambda^{2t-1}x$ and define
$\by_0=(0,\ldots,0,1)$, and
$$\by_i={\sqrt{1-\beta^2 \over q(1)}}
\left(\underbrace{\bu_i' \otimes \ldots \otimes
\bu_i'}_{\mbox{$2t$ times}},
\sqrt{2t\lambda^{2t-1}}\bu_i',0\right) + \beta \by_0,$$
where $\beta$ will be determined later. Note that $\by_i$ is
normalized to satisfy $\|\by_i\|=1$.

Moreover $\by_i$ is defined so that $\by_i\cdot \by_j$ takes its
minimum value when $ij \in E$, i.e., when $\bu_i'\cdot
\bu_j'=-\lambda$. As is shown in \cite{Char02}, for every $\eps>0$
we may set $t = \Omega({1\over \eps}), \beta=\Theta(1/t), \gamma =
{1\over 4t}$ to get that $(\by_0-\by_i)\cdot (\by_0-\by_j)=0$ for
$ij \in E$, while $(\by_0-\by_i)\cdot (\by_0-\by_j)\ge 0$ always.


Now we verify that all the triangle inequalities, i.e., the second
constraint of SDP (\ref{SDP-VC2}) are satisfied: First note that
since every coordinate takes only two different values for the
vectors in $\{\by_i: i \in V\}$, it is easy to see that
$c_1(\{\by_i: i \in V\},\|\cdot\|^2)=1$. So the triangle inequality
holds when $i,j,k \in V$. When $i=0$ or $j=0$, the inequality is
trivial, and it only remains to verify the case that $k=0$, i.e.,
$(\by_0-\by_i)\cdot (\by_0-\by_j) \ge 0$, which was already
mentioned above. Now
$ \sum_{i \in V} (1+\by_0\cdot \by_i)/2 =\frac{1+\beta}{2} \cdot |V|
=\left(\frac{1}{2}+O(\eps) \right)|V|$,
where by the result of Frankl and R\"{o}dl $\vc(G)=(1-o(1))|V|$.

\subsection{Karakostas' and Pentagonal SDP formulations}
\label{Sec:strongerSDP} 
Karakostas suggests the following SDP relaxation, that is the result of 
adding to SDP (\ref{SDP-VC3}) the triangle inequalities applied to the 
set  $\{\bv_i : i \in V \cup \{0\}\} \cup \{-\bv_i : i \in V\cup \{0\}\} $.
\begin{equation}
\label{SDP-Karakostas}
\begin{array}{clcl}
\Min & \sum_{i \in V} (1+\bv_0\bv_i)/2 &\qquad &\\
{\rm s.t.} &  (\bv_i-\bv_0)\cdot (\bv_j-\bv_0) = 0  & &\forall \ ij \in E \\
&  (\bv_i - \bv_k) \cdot (\bv_j - \bv_k) \ge 0  & &\forall \ i,j,k \in V \\
&  (\bv_i + \bv_k) \cdot (\bv_j - \bv_k) \ge 0  & &\forall \ i,j,k \in V \\
&  (\bv_i + \bv_k) \cdot (\bv_j + \bv_k) \ge 0  & &\forall \ i,j,k \in V \\
&\| \bv_i \| = 1  & & \forall \ i \in \{0\} \cup V.
\end{array}
\end{equation}
We prove that this variant has integrality gap $2-o(1)$ by
showing that Charikar's construction satisfies SDP
(\ref{SDP-Karakostas}). We postpone the proof to the appendix.

\begin{theorem}
\label{Thm:Karakostas} The integrality gap of SDP
(\ref{SDP-Karakostas}) is bigger than $2-\eps$, for any $\eps
>0$.
\end{theorem}

By now we know that taking all the $\ell_1$ constraints leads to
an exact relaxation, but clearly one that is not tractable. Our goal 
here is to explore the possibility that stepping towards $\ell_1$ 
embeddability while still maintaining computational feasibility 
would considerably reduce the integrality gap. A canonical set of valid
inequalities for $\ell_1$ metrics is the so called {\em
Hypermetric inequalities}. Metrics that satisfy all these
inequalities are called {\em hypermetrics}. Again, taking all
these constraints is not feasible, and yet we do not know whether
this may lead to a better integrality gap (notice that we do not
know that Theorem \ref{Thm:SDP-l1} remains true if we replace the
$\ell_1$ embeddability constraints with a hypermetricity
constraint). See \cite{dela97} for a related discussion about
hypermetrics. We instead consider the effect of adding a small number 
of such constraints. The simplest hypermetric inequalities 
beside triangle inequalities are the {\em pentagonal} inequalities. 
These inequalities consider two sets of points in the space of
size 2 and 3, and require that the sum of the distances between points
in different sets is at least the sum of the distances within sets.
Formally, let $S,T \subset X$, $|S|=2,|T|=3$, then we have
the inequality
$\sum_{i\in S,j\in T}d(i,j) \ge \sum_{i,j\in S}d(i,j) + \sum_{i,j\in T}d(i,j)$.
To appreciate this inequality it is useful to describe where it fails. 
Consider the graph metric of $K_{2,3}$. Here, the LHS of the inequality is 6 
and the RHS is 8, hence $K_{2,3}$ violates the pentagonal inequality. 
In the following theorem we show that this ``next level'' strengthening past 
the triangle inequalities fails to reduce the integrality gap significantly.

\begin{theorem}
\label{Thm:Pentagonal} The integrality gap of SDP (\ref{SDP-VC4}) is at
least $2-\eps$ for any $\eps >0$.
\end{theorem}
We give here an outline of the proof (the complete proof appears in the appendix).

We resort again to Charikar's construction. Recall that by ignoring 
$\by_0$ the metric space defined by $d(i,j) = \|\by_i-\by_j\|^2$ is $\ell_1$ embeddable. 
Therefore, the only $\ell_1$-valid inequalities that may be violated are ones
containing $\by_0$. Hence, we wish to consider a pentagonal inequality 
containing $\by_0$ and four other vectors, denoted by 
$\by_1,\by_2,\by_3,\by_4$. Assume first that the partition of the five points 
in the inequality puts ${\by_0}$ together with two other points; then, using
the fact that 
$d(0,1) = d(0,2) = d(0,3) = d(0,4)$ and triangle inequality we get that
such an inequality must hold.
It therefore remains to consider a partition of the form 
$(\{\by_1,\by_2,\by_3\},\{\by_4,\by_0\})$, in other words we need
to show that: $$d(1,2)+ d(1,3) +d(2,3) + d(0,4) \le d(1,4)
+d(2,4)+d(3,4)+d(0,1)+d(0,2)+d(0,3)$$

Let $q(x) = x^{2t}+2t\lambda^{2t-1}x$. Recall that every $\by_i$ is associated with a $\{-1,1\}^n$ unit vector $\bu_i$ and its scaled counterpart $\bu'_i$. After substituting each 
$\by_i$ as a function of $\bu'_i$, the inequality gets the form 
\begin{equation}
\label{ineq2} 
E = q(\bu'_1\cdot \bu'_2)+q(\bu'_1\cdot
\bu'_3)+q(\bu'_2\cdot \bu'_3)-q(\bu'_1\cdot \bu'_4)-q(\bu'_2\cdot
\bu'_4)-q(\bu'_3\cdot \bu'_4) \ge-2q(1)/(1+\beta)
\end{equation}

The rest of the proof analyzes the minima of the function $E$ and ensures
that (\ref{ineq2}) is satisfied at those minima. 
We proceed by first partitioning the coordinates of the original hypercube 
into four sets according to the sign of $\bu_1,\bu_2$ and $\bu_3$ on 
these coordinates. We let $P_0$ be the set of coordinates in which all 
three vectors assume negative value, and $P_1 (P_2,P_3)$ be the 
coordinates on which $\bu_1 (\bu_2,\bu_3)$ is positive and the other two 
vectors negative. Without loss of generality the union of these four sets is the set 
of all coordinates.
Next, $\bu_4$ is considered. Using the convexity of the polynomial $q$ we 
show that we may assume that $\bu_4$ is either all $1$ or all $-1$ on each 
set $P_i$. 
Stronger properties of $q$ ensure that $\bu_4$ is $-1$ on the $P_0$ 
coordinates.

The cases left to check now are characterized by whether $\bu_4$
is $1$ or $-1$ on each of $P_1,P_2,P_3$. By symmetry, all we
need to know is the number of blocks $P_i$ on which $\bu_4$ takes the value $1$. Hence we are left with four cases and we use calculus arguments to analyze each case separately. Our analysis shows that in all cases the function $E$ is minimized when 
$\bu_4$ identifies with one of $\bu_1,\bu_2,\bu_3$; but then it can be easily 
seen that the pentagonal inequality reduces to a triangle inequality which we
know is valid. 

\section{Lower bound for embedding negative type metrics into $\ell_1$} \label{sec:l2metl1}
While, in view of Theorem \ref{Thm:SDP-weakl1}, Charikar's metric
does not supply an example that is far from $\ell_1$, we may still
(partly motivated by Theorem \ref{Thm:SDP-l1}) utilize the idea of
``tensoring the cube'' and then adding some more points in order
to achieve negative type metrics that are not $\ell_1$ embeddable.
Our starting point is an isoperimetric inequality on the cube that
generalizes the standard one, and under certain conditions provides
better edge expansion guarantee. Such a setting is also relevant in
\cite{KhotVishnoi,KrauthgamerRabani} where harmonic analysis tools
are used to bound expansion; there tools are unlikely to be applicable to our
case where the interest and improvements lie in the constants.

\begin{theoremone}
(Generalized Isoperimetric inequality) For every
set $S \subseteq Q_n$,
$$|E(S,S^c)| \ge |S|(n-\log_2 |S|)+p(S).$$
where $p(S)$ denotes the number of
vertices $\bu \in S$ such that $-\bu \in S$. 
\end{theoremone}
\begin{proof}
We use induction on $n$. Divide $Q_n$ into two sets
$V_{1}=\{\bu:\bu_1=1\}$ and $V_{-1}=\{\bu:\bu_1=-1\}$. Let
$S_{1}=S \cap V_{1}$ and $S_{-1}=S \cap V_{-1}$. Now, $E(S,S^c)$
is the disjoint union of $E(S_1,V_1 \setminus S_1)$,
$E(S_{-1},V_{-1} \setminus S_{-1})$, and $E(S_1,V_{-1} \setminus
S_{-1}) \cup E(S_{-1},V_{1} \setminus S_1)$. Define the operator
$\widehat{\cdot}$ on $Q_n$ to be the projection onto the last
$n-1$ coordinates, so for example $\widehat{S_1}=\{\bu \in
Q_{n-1}: (1,\bu) \in S_1\}$.
It is easy to observe that
$$|E(S_1,V_{-1}\setminus S_{-1}) \cup
E(S_{-1},V_{1} \setminus S_1)|=|\widehat{S_1} \Delta
\widehat{S_{-1}}|.$$
We now argue that
\begin{equation}
\label{divide} p(S) + |S_1| - |S_{-1}| \le
p(\widehat{S_1})+p(\widehat{S_{-1}})+|\widehat{S_1} \Delta
\widehat{S_{-1}}|.
\end{equation}
To prove (\ref{divide}), for every $\bu \in \{-1,1\}^{n-1}$, we
show that the contribution of $(1,\bu)$, $(1,-\bu)$, $(-1,\bu)$,
and $(-1,-\bu)$ to the right hand side of (\ref{divide}) is at
least as large as their contribution to the left hand side: This
is trivial if the contribution of these four vectors to $p(S)$ is
not more than their contribution to $p(\widehat{S_{1}})$, and
$p(\widehat{S_{-1}})$. We therefore assume that the contribution
of the four vectors to $p(S)$, $p(\widehat{S_{1}})$, and
$p(\widehat{S_{-1}})$ are $2$, $0$, and $0$, respectively. Then
without loss of generality we may assume that $(1,\bu),(-1,-\bu)
\in S$ and $(1,-\bu),(-1,\bu) \not\in S$, and in this case the
contribution to both sides is $2$.

By induction hypothesis and (\ref{divide}) we get
\begin{eqnarray*}
|E(S,S^c)| &=&|E(\widehat{S_1},Q_{n-1} \setminus \widehat{S_1}|
+             |E(\widehat{S_{-1}},Q_{n-1} \setminus \widehat{S_{-1}}|
+             |\widehat{S_1} \Delta \widehat{S_{-1}}|
\\ &\ge &
|S_1|(n-1-\log_2 |S_1|)+p(\widehat{S_1})+|S_{-1}|(n-1-\log_2 |S_{-1}|)+
p(\widehat{S_{-1}}) +|\widehat{S_1} \Delta \widehat{S_{-1}}|\\
& \ge & |S|n-|S|-(|S_1|\log_2 |S_1|+|S_{-1}|\log_2 |S_{-1}|)
+p(\widehat{S_1})+p(\widehat{S_{-1}})+|\widehat{S_1} \Delta \widehat{S_{-1}}| \\
& \ge & |S|n-(2|S_{-1}|+|S_1|\log_2 |S_1|+|S_{-1}|\log_2 |S_{-1}|)
+ p(S).
\end{eqnarray*}
Now the lemma follows from the fact that
%
$2|S_{-1}|+|S_1|\log_2 |S_1|+|S_{-1}|\log_2 |S_{-1}| \le |S| \log_2 |S|$,
which can be obtained using easy calculus.
\end{proof}

We call a set $S
\subseteq Q_n$ {\em symmetric} if $-\bu \in S$ whenever $\bu \in
S$. Note that $p(S)=|S|$ for symmetric sets $S$.
\begin{corollary}
For every symmetric set $S \subseteq Q_n$
$$|E(S,S^c)| \ge |S|(n-\log_2 |S|+1).$$
\end{corollary}

The corollary above implies the following Poincar\'{e}
inequality.

\begin{proposition} \label{Poincare}
(Poincar\'{e} inequality for the cube and an additional point)
Let $f:Q_n \cup \{{\bf 0}\} \rightarrow \R^m$ satisfy that
$f(\bu)=f(-\bu)$ for every $\bu \in Q_n$.
Then the following Poincar\'{e} inequality holds.
\begin{equation}
\label{poincareEq} {1\over  2^n}\cdot {8\over 7}(4 \alpha+1/2)
\sum_{\bu,\bv \in Q_n}\|f(\bu)-f(\bv)\|_1 \le \alpha\sum_{\bu\bv \in E}
\|f(\bu)-f(\bv)\|_1 + \frac{1}{2}\sum_{\bu\in Q_n}\|f(\bu)-f({\bf
0})\|_1
\end{equation}
where $\alpha=\frac{\ln 2}{14 - 8\ln 2}.$
\end{proposition}
\begin{proof}
It is well known that instead of considering $f:V\rightarrow
\ell_1$, it is enough to prove the above inequality for
$f:V\rightarrow\{0,1\}$. Further, we may assume without loss of
generality that $f({\bf 0}) = 0$. Associating $S$ with
$\{\bu:f(\bu)=1\}$, Inequality (\ref{poincareEq}) reduces to
\begin{equation}\label{boolpoinc}
{1\over  2^n}{8\over 7}(4 \alpha+1/2) |S||S^c| \le \alpha
|E(S,S^c)|  + |S|/2,
\end{equation}
where $S$ is a symmetric set, owing to the condition $f(\bu) =
f(-\bu)$. From the isoperimetric inequality of
Theorem~\ref{Thm:isoper} we have that $|E(S,S^c)| \ge |S|(x+1)$
for $x= n - \log_2|S|$ and so
$$\left(\frac{\alpha (x+1)+1/2}{1-2^{-x}}\right) \frac{1}{2^n}
|S||S^c)| \le \alpha |E(S,S^c)|  + |S|/2.$$ It can be verified
(See Lemma~\ref{calculus}) that $\frac{\alpha
(x+1)+1/2}{1-2^{-x}}$ attains its minimum in $[1,\infty)$ at $x=3$
whence  $\frac{\alpha (x+1)+1/2}{1-2^{-x}} \ge {4\alpha+1/2 \over
7/8}$, and Inequality (\ref{boolpoinc}) is proven.
\end{proof}

\begin{theorem}
\label{Thm:distortion} Let $V=\{\tilde{\bu}: \bu \in Q_n\} \cup
\{{\bf 0}\}$, where $\tilde{\bu}=\bu \otimes \bu$. Then for the
semi-metric space $X=(V, \|\cdot\|^2)$ we have $c_1(X) \ge
\frac{8}{7}-\eps$, for every $\eps>0$ and sufficiently large $n$.
\end{theorem}
\begin{proof}
We start with an informal description of the proof. The heart of the argument 
is showing that the cuts that participate in a
supposedly good $\ell_1$ embedding of $X$ cannot be balanced on one hand, 
and cannot be imbalanced on the other.
First notice that the average distance in $X$ is almost double that of the distance between $\bo$ and any other 
point (achieving this in a cube structure without violating the triangle 
inequality was where the tensor operation came in handy). 
For a cut metric on the points of $X$, such a relation
only occurs for very imbalanced cuts; hence the representation of balanced 
cuts in a low distortion embedding cannot be large. 
On the other hand,
comparing the (overall) average distance to the average distance between 
neighbouring points in the cube shows that any good embedding must use cuts with 
very small edge expansion, and such cuts in the cube must be balanced 
(the same argument says that one must use the dimension cuts when embedding 
the hamming cube into $\ell_1$ with low distortion). The fact that only 
{\em symmetric cuts} participate in the $\ell_1$ embedding (or else the 
distortion becomes infinite due to the tensor operation) enables us to use 
the stronger isoperimetric inequality which leads to the current lower bound. 
We proceed to the proof itself.

We may view $X$ as a distance function with points in $ \bu \in
Q_n \cup \{{\bf 0}\}$, and $d(\bu,\bv) = \|\tilde{\bu}  -
\tilde{\bv}\|^2$. We first notice that $X$ is indeed a metric
space, i.e., that triangle inequalities are satisfied: notice that
$X \setminus \{\bo\}$ is a subset of $\{-1,1\}^{n^2}$. Therefore, the
square Euclidean distances is the same (upto a constant) as their
$\ell_1$ distance. Hence, the only triangle inequality we need to check is
$\|\tilde{\bu}-\tilde{\bv}\|^2 \le \|\tilde{\bu}-{\bf 0}\|^2 +
\|\tilde{\bv}-{\bf 0}\|^2$, which is implied by the fact that
$\tilde{\bu}\cdot \tilde{\bv} =(\bu\cdot \bv)^2$ is always
nonnegative.

For every $\bu,\bv \in Q_n$, we have $d(\bu,{\bf 0}) =
\|\tilde{\bu}\|^2 = \tilde{\bu} \cdot \tilde{\bu} = (\bu\cdot
\bu)^2=n^2$, and $d(\bu,\bv) =
\|\tilde{\bu}-\tilde{\bv}\|^2=\|\tilde{\bu}\|^2+\|\tilde{\bv}\|^2-2(\tilde{\bu}
\cdot \tilde{\bv})  = 2n^2-2(\bu\cdot \bv)^2$. In particular, if
$\bu \bv \in E$ we have $d(\bu,\bv) = 2n^2-2(n-2)^2 = 8(n-1)$. We
next notice that
$$\sum_{\bu,\bv \in Q_n}d(\bu,\bv)=
2^{2n}\times 2n^2-2\sum_{\bu,\bv}(\bu \cdot \bv)^2=
2^{2n}\times 2n^2-2\sum_{\bu,\bv}(\sum_i \bu_i\bv_i)^2
=2^{2n}(2n^2-2n),$$
as $\sum_{\bu,\bv} \bu_i\bv_i\bu_j\bv_j$ is $2^{2n}$ when $i=j$,
and $0$ otherwise.

Let $f$ be a nonexpanding embedding of $X$ into $\ell_1$. Notice
that $$d(\bu,-\bu) = 2n^2-2(\bu\cdot \bv)^2 = 0,$$ and so any
embedding with finite distortion must satisfy $f(\bu)=f(-\bu)$.
Therefore Inequality (\ref{poincareEq}) can be used and we get
that
\begin{equation}
\label{poincfrac} \frac{\alpha \sum_{\bu\bv \in
E}\|f(\tilde{\bu})-f(\tilde{\bv})\|_1+ \frac{1}{2}\sum_{\bu \in
Q_n} \|f(\tilde{\bu})-f({\bf 0})\|_1} {\frac{1}{2^n}\sum_{\bu,\bv
\in Q_n}\|f(\tilde{\bu})-f(\tilde{\bv})\|_1} \ge {8\over 7}(4
\alpha+1/2).
\end{equation}

On the other hand,
\begin{equation}
\label{fractionOriginal} \frac {\alpha \sum_{\bu\bv \in
E}d(\bu,\bv) + {1 \over 2}\sum_{\bu\in Q_n} d(\bu,{\bf 0})}{{1
\over 2^n}\sum_{\bu,\bv \in Q_n}d(\bu,\bv)}=
\frac{8\alpha(n^2-n)+n^2}{2n^2-2n}= 4 \alpha+1/2 + o(1).
\end{equation}

The discrepancy between (\ref{poincfrac}) and
(\ref{fractionOriginal}) shows that for every $\eps>0$ and
for sufficiently large $n$, the required distortion of $V$ into
$\ell_1$ is at least $8/7-\eps$.
\end{proof}

\section{Conclusion}
We have considered the metric characterization of SDP relaxations
 of
{\sc Vertex Cover} and specifically related the amount of ``$\ell_1$ 
information'' that is enforced with the resulting integrality gap. We
showed 
 that a $2-o(1)$ integrality gap survives in the feasible
extreme of this 
 range, while no integrality gap exists in the most
powerful (and not feasible)
 extreme, i.e., when $\ell_1$
embeddability of the solution is
 enforced. We further demonstrated
that integrality gap is not a
 continuous function of the possible
distortion that is allowed,
 as it jumps from 1 to $2-o(1)$ when the
allowed distortion changes from 
 1 to $1+\delta$. These results
motivated us to find a negative type metric 
 that does not embed well
to $\ell_1$, which is a fairly elusive object. 
 The natural
extensions of these results are to (i) check whether the addition of
more $k$-gonal inequalities
 (something that can be done efficiently
for any finite number of
 such inequalities) can reduce the
integrality gap or prove
 otherwise. We in fact conjecture that the
integrality gap is still $2-o(1)$
 when we impose the condition that
the solution is a Hypermetric. It is interesting to note that related
questions are discussed in the context of LP relaxations of {\sc
Vertex Cover} in \cite{ABLT05} (ii) use 
 the nonembeddability
construction and
 technique in Section \ref{sec:l2metl1} to find
negative type
 metrics that incur more significant distortion when
embedded into
 $\ell_1$. It is interesting to investigate whether (and
how) our
 findings are connected to the question of the power of Lift
and
 Project methods; specifically the one that is defined with the
Positive Semi Definiteness constraints, also known as
 $\mbox{LS}^+$
(see \cite{AAT05} for relevant discussion). Notice
 that $k$ rounds of
LS$^+$ will imply all $k$-gonal inequalities,
 but may be much
stronger. In fact, we do not even know whether
 applying two rounds of
LS$^+$ does not lead to an integrality gap
 of $2-\Omega(1)$.
 Last,
we suggest looking at connections of $\ell_1$-embeddability and 
integrality gaps for other NP-hard problems. Under certain
circumstances, such 
 connections may be used to convert hardness
results of combinatorial problems
 into hardness results of
approximating $\ell_1$ distortion.
 
\section*{Acknowledgment} Special thanks to George Karakostas for
very valuable discussions.


\bibliographystyle{plain}
\bibliography{tensor}



\section{Appendix}

\subsection{Proof of Theorem~\ref{Thm:SDP-weakl1} \label{Sec:proof-SDP-weakl1}}
Let $\by_i$ and $\bu_i'$ be defined as in Section~\ref{Charikar}.
To prove Theorem~\ref{Thm:SDP-weakl1}, it is sufficient to prove
that $c_1(\{\by_i:i \in \{0\} \cup V\},\|\cdot\|^2)=1+o(1)$. Note that
every coordinate of $\by_i$ for all $i \in V$ takes at most two
different values. It is easy to see that this implies
$c_1(\{\by_i:i \in  V\},\|\cdot\|^2)= 1$. In fact
\begin{equation}
f:\by_i \mapsto {1-\beta^2 \over q(1)}
\left(\frac{2}{n^t}\underbrace{\bu_i' \otimes \ldots \otimes
\bu_i'}_{\mbox{$2t$
times}},\frac{2}{\sqrt{n}}2t\lambda^{2t-1}\bu_i'\right),
\end{equation}
is an isometry from $(\{\by_i:i \in  V\},\|\cdot\|^2)$ to
$\ell_1$. For $i \in V$, we have
\begin{equation}
\|f(\by_i)\|_1 = {1-\beta^2 \over q(1)} \left(\frac{2}{n^t} \times
\frac{n^{2t}}{n^t}+\frac{2}{\sqrt{n}}2t\lambda^{2t-1}\frac{1}{\sqrt{n}}+0\right)
={1-\beta^2 \over q(1)}\times(2+4t\lambda^{2t-1})
\end{equation}
Since $\beta=\Theta(\frac{1}{t})$, recalling that
$\lambda=1-\frac{1}{2t}$, it is easy to see that for every $i\in
V$, $\lim_{t \rightarrow \infty} \|f(\by_i)\|_1=2$. On the other
hand for every $i \in V$
$$\lim_{t \rightarrow \infty}\|\by_i-\by_0\|^2=\lim_{t \rightarrow \infty}2-2(\by_i\cdot \by_0)=
\lim_{t\rightarrow \infty}2-2\beta=2.$$
So if we extend $f$ to $\{\by_i: i \in V \cup \{0\}\}$ by defining
$f(\by_0)={\bf 0}$, we obtain a mapping from $(\{\by_i: i \in V
\cup \{0\}\}, \|\cdot\|^2)$ to $\ell_1$ whose distortion tends to $1$
as $t$ goes to infinity

%

\subsection{Proof of Theorem~\ref{Thm:Karakostas}}
We show that the Charikar's construction satisfies formulation
(\ref{SDP-Karakostas}). By~\cite{Char02} and from the discussion
in Section~\ref{Charikar}, it follows that all edge constraints
and triangle inequalities of the original points hold. Hence we
need only consider triangle inequalities with at least one
nonoriginal point. By homogeneity, we may assume that there is exactly one
such point.

Since all coordinates of $\by_i$ for $i>0$ assume only two values
with the same absolute value, it is clear that not only does the
metric they induce is $\ell_1$ but also taking $\pm \by_i$ for
$i>0$ gives an $\ell_1$ metric; in particular all triangle
inequalities that involve these vectors are satisfied. In fact, we
may fix our attention to triangles in which $\pm \by_0$ is the
middle point. This is since
$$(\pm \by_i - \pm \by_j) \cdot (\by_0 - \pm \by_j) =
(\pm \by_j - \by_0) \cdot (\mp \by_i - \by_0).$$

Consequently, and using symmetry, we are left with checking the
nonnegativity of $(\by_i + \by_0) \cdot (\by_j + \by_0)$ and
$(-\by_i - \by_0)\cdot (\by_j  - \by_0)$.
$$(\by_i + \by_0) \cdot (\by_j + \by_0) = 1 + \by_0 \cdot (\by_i+\by_j)+\by_i \cdot \by_j \ge
1 + 2\beta + \beta^2 - (1-\beta^2) = 2\beta(1+\beta) \ge 0.$$
Finally,
$$(-\by_i - \by_0)\cdot (\by_j  - \by_0) =
1 +\by_0 \cdot (\by_i-\by_j) -\by_i \cdot \by_j = 1-\by_i \cdot
\by_j \ge 0
$$ as $\by_i,\by_j$ are of norm 1.

\subsection{Proof of Theorem~\ref{Thm:Pentagonal}}
Again we show that the metric space used in Charikar's
construction satisfies the pentagonal inequalities. As explained in the outline of the proof in Section \ref{Sec:strongerSDP}, we need to consider only pentagonal inequalities in which the partition of the vectors is of the form $(\{\by_1,\by_2,\by_3\}, \{\by_4,\by_0 \})$.
Therefore we need
to show that: $$d(1,2)+ d(1,3) +d(2,3) + d(0,4) \le d(1,4)
+d(2,4)+d(3,4)+d(0,1)+d(0,2)+d(0,3)$$



As the vectors are of unit norm, it is clear that $d(0,i) = 2-2\beta$ for all $i>0$ and that $d(i,j) = 2-2\by_i\by_j$.
Recall that every $\by_i$ is associated with a $\{-1,1\}$ vector
$\bu_i$ and with its normalized multiple $\bu'_i$. Also, it is
simple to check that $\by_i\cdot \by_j =
\beta^2+(1-\beta^2)q(\bu'_i\cdot \bu'_j)/q(1)$ where $q(x) =
x^{2t}+2\lambda^{2t-1}x$. After substituting this in the previous
expression, it is easy to see that our goal is then to show:
\begin{equation}
\label{ineq.} E = q(\bu'_1\cdot \bu'_2)+q(\bu'_1\cdot
\bu'_3)+q(\bu'_2\cdot \bu'_3)-q(\bu'_1\cdot \bu'_4)-q(\bu'_2\cdot
\bu'_4)-q(\bu'_3\cdot \bu'_4) \ge-2q(1)/(1+\beta)
\end{equation}


We partition the coordinates of the original hypercube into four
sets according to the values assumed by $\bu_1,\bu_2$ and $\bu_3$.
Assume without loss of generality that in any coordinate at most
one of these get the value -1 (otherwise multiply the values of
the coordinate by $-1$). We get four sets, $P_0$ for the
coordinates in which all three vectors assume value -1, and
$P_1,P_2,P_3$ for the coordinates in which exactly
$\bu_1,\bu_2,\bu_3$ respectively assumes value 1.

We now consider $\bu_4$. We argue that without loss of generality
we may assume that $\bu_4$ is ``pure'' on each of the
$P_0,P_1,P_2,P_3$; in other words it is either all 1 or all $-1$ on
each one of the them. Assume for sake of contradiction that there
are $w$ coordinates in $P_0$ on which $\bu_4$ assumes value $-1$,
and that $0<w<|P_0|$. Let $\bu_4^+$ (similarly $\bu_4^-$) be
identical to $\bu_4$ except we replace one 1 in $P_0$ by $-1$
(replace one $-1$ in $P_0$ by 1). We show that replacing $\bu_4$
by $\bu_4^+$ or by $\bu_4^-$ we decrease the expression $E$. This
means that the original $\bu_4$ could not have been a choice that
minimized $E$ and the claim follows. Let $p_i = \bu_i \cdot
\bu_4$, $p_i^+ = \bu_i' \cdot (\bu^+_4)'$ and $p_i^- = \bu'_i
\cdot (\bu^-_4)'$ for $i = 1,2,3$. Notice that the above
replacement only changes the negative terms in~(\ref{ineq.}) so
our goal now is to show that $\sum_{i=1}^3 q(p_i) <
\max\{\sum_{i=1}^3 q(p_i^+) , \sum_{i=1}^3 q(p_i^-)\}$.
$$\max\{\sum_{i=1}^3 q(p_i^+) , \sum_{i=1}^3 q(p_i^-)\} \ge
{\sum_{i=1}^3 q(p_i^+) + \sum_{i=1}^3 q(p_i^-) \over 2} =
$$
$$\sum_{i=1}^3 {q(p_i^+) + q(p_i^-) \over 2} >
\sum_{i=1}^3 q\left({p_i^+ + p_i^- \over 2}\right) = \sum_{i=1}^3
q(p_i),$$ where the second last inequality is using the (strict)
convexity of $q$. This of course applies to $P_1,P_2$ and $P_3$ in
precisely the same manner. The above characterization
significantly limits the type of configurations we need to check
but regretfully, there are still quite a lot of cases to check. 

For $P_0$, we can in fact say something stronger than we do for
$P_1,P_2,P_3$:
\begin{proposition}
\label{Prop:w}
If there is a violating configuration, there is one with $\bu_4$ that has all the $P_0$ coordinates set to 
$-1$.
\end{proposition}
This is not a surprising fact; in fact if $q$ was a monotone
increasing function this would be obvious, but of course the whole
point behind $q$ is that it brings to minimum some intermediate value
($-\lambda$) and hence can not be increasing. The convexity of $q$ is
also not enough, and one should really utilize the exact properties of
$q$.  We postpone the proof till the end and continue our analysis
assuming the proposition.

The cases left to check now are characterized by whether $\bu_4$ is
$1$ or $-1$ on each of $P_1,P_2,P_3$. By symmetry all we really need
to know is $$\xi(\bu_4) = |\{i ~:~ \bu_4 \mbox { is 1 on }P_i\}|$$ If
$\xi(\bu_4)=1$ it means that $\bu_4$ is the same as one of
$\bu_1,\bu_2$ or $\bu_4$ hence the pentagonal inequality reduces to
the triangle inequality, which we have already shown is valid. If
$\xi(\bu_4)=3$, it is easy to see that in this case $\bu'_1 \bu'_4 =
\bu'_2 \bu'_3$, and likewise $\bu'_2 \bu'_4 = \bu'_1 \bu'_3$ and
$\bu'_3 \bu'_4 = \bu'_1
\bu'_2$ hence $E$ is $0$ for these cases, which means that the 
inequality~\ref{ineq.} is satisfied.

We are left with the cases $\xi(\bu_4) \in \{0,2\}$. 

\noindent {\bf Case 1:} $\xi(\bu_4) = 0$

Let $x = {2 \over n}|P_1|,y = {2 \over n}|P_2|, z = {2 \over
n}|P_3|$. Notice that $x+y+z  = {2 \over n}(|P_1|+|P_2|+|P_3|) \le
2$, as these sets disjoint. Now, think of
$$E = E(x,y,z) =q(1-(x+y)) +q(1-(x+z)) + q(1-(y+z)) -q(1-x) -q(1-y) -q(1-z)$$
as a function from $\R^3$ to $\R$, and we will show the (stronger
than necessary) claim that $E$ achieves its minimum in
$\{(x,y,z)\in \R^3 : x+y+z \le 2\}$ at points where either $x,y$
or $z$ are zero. Assume without loss of generality that $0\leq x\leq y
\leq z$.

We consider the function $g(\delta) = E(x-\delta,y+\delta,z)$. It is easy to see 
that $g'(0) =
q'(1-(x+z))-q'(1-(y+z))-q'(1-x)+q'(1-y)$. Our goal is to show that
$g'(0)$ is nonpositive, and in fact that $g'(\delta) \le 0$ for
every $\delta \in [0,x]$. This, by the Mean Value Theorem implies that
$$E(0,x+y,z) \le E(x,y,z),$$ and in particular that in this case
we may assume that $x=0$. This means that $\by_1 = \by_4$ which reduces to
the triangle inequality on $\by_0,\by_2,\by_3$.

Note that in $q'(1-(x+z))-q'(1-(y+z))-q'(1-x)+q'(1-y)$, the two
arguments in the terms with positive sign have the same average as
the arguments in the terms with negative sign, namely
$\mu = 1-(x+y+z)/2$. We now have $g'(0) =
q'(\mu+b)-q'(\mu+s)-q'(\mu-s)+q'(\mu-b)$, where $b = (x-y+z)/2, s
= (-x+y+z)/2$.
\begin{eqnarray*}
g'(0) & = &[q'(\mu+b)+q'(\mu-b) - q'(\mu+s)-q'(\mu-s)]\\
& = & 2t[(\mu+b)^{2t-1} + (\mu-b)^{2t-1} -
(\mu+s)^{2t-1} - (\mu-s)^{2t-1}]\\
& = & 4t\sum_{i \mbox{ even}} {2t-1 \choose i} \mu^{2t-1-i}(b^i -
s^i)
\end{eqnarray*}

Notice that $\mu = 1-(x+y+z)/2 \ge 0$. Further, since $x\le y$, we get
that $s \ge b\ge 0$. This means that $g'(0) \le 0$. It can be easily
checked that the same argument holds if we replace $x,y$ by $x-\delta$
and $y+\delta$. Hence $g'(\delta) \le 0$ for every $\delta \in [0,x]$,
and we are done.

\noindent {\bf Case 2:} $\xi(\bu_4) = 2$

The expression for $E$ is now:
$$ E(x,y,z) = q(1-(x+y))+q(1-(x+z))+q(1-(y+z))-q(1-x)-q(1-y)-q(1-(x+y+z))$$

Although $E(x,y,z)$ is different than in Case 1, the important observation is that if we consider again the function $g(\delta) = E(x-\delta, y+\delta,z)$ then the derivative $g'(\delta)$ is the same as in Case 1 and hence the same analysis shows that $E(0,x+y,z)\le E(x,y,z)$. Therefore we may assume that $x=0$. This means that $\by_2$ identifies with $\by_4$ and the inequality reduces to
the triangle inequality on $\by_0,\by_1,\by_3$.

\noindent It now remains to prove Proposition \ref{Prop:w}:

\begin{proofof}{Proof of Proposition \ref{Prop:w}}
Fix a configuration for $\bu_1,\bu_2,\bu_3$ and as before let $x = \frac{2}{n}|P_1|$, $y = \frac{2}{n}|P_2|$, $z = \frac{2}{n}|P_3|$, and $w = \frac{2}{n}|P_0|$,
where $w>0$. Consider a vector $\bu_4$ that has all
$-1$'s in $P_0$. Let $H_i = \frac{2}{n}H(\bu_i,\bu_4)$, where
$H(\bu_i,\bu_4)$ is the Hamming distance from $\bu_4$ to $\bu_i$, $i=1,2,3$. 
It suffices to show that replacing
the $P_0$-part of $\bu_4$ with $1$'s (which means adding $w$ to each $H_i$) 
does not decrease the LHS of~\ref{ineq.},
i.e., that:
\begin{equation}
\label{Eq:w}
q(1-H_1) + q(1-H_2)+ q(1-H_3) \geq
q(1-(H_1+w)) +q(1-(H_2+w))+q(1-(H_3+w))
\end{equation}

Because of the convexity of $q$ as explained before, the cases that we need to consider are characterized by whether $\bu_4$ is $1$ or $-1$ on each of $P_1,P_2,P_3$. By symmetry there are $4$ cases to check, corresponding to the different values of $\xi(\bu_4)$. In some of these cases, we use the following argument: consider the function $g(\delta) = q(1-(H_1+\delta)) +q(1-(H_2+\delta))+q(1-(H_3+\delta))$, where $\delta\in [0,w]$. 
Let $a_i = 1- (H_i+\delta)$. The derivative $g'(\delta)$ is:
$$  g'(\delta) = -(q'(a_1)+q'(a_2)+q'(a_3)) =
-2t(a_1^{2t-1}+a_2^{2t-1}+a_3^{2t-1}+3\lambda^{2t-1})$$
If we show that the derivative is negative for any $\delta\in [0,w]$, that would
imply that $g(0)\geq g(w)$ and hence we are done since we have a more violating configuration if we do not add
$w$ to the Hamming distances. 

\noindent {\bf Case 1:} $\xi(\bu_4) = 0$

In this case $H_1 = x$, $H_2 = y$, $H_3 = z$.
Note that $x+y+z+w = 2$. Hence, if $H_i\geq 1$ for some $i$, say for $H_1$, then $H_2+\delta\leq 1$ and $H_3+\delta\leq 1$. This implies that $a_2\geq 0$ and $a_3\geq 0$. Thus 
$$ g'(\delta)\leq -( -1 +3\lambda^{2t-1}) \leq 1-3/e<0$$
since $\lambda^{2t-1} = (1-\frac{1}{2t})^{2t-1}\geq 1/e$. Hence we are done.

Therefore, we can assume that $H_i< 1$ for all $i$, i.e., $1-H_i\geq 0$. We now compare the LHS and RHS of (\ref{Eq:w}). In particular we claim that each term $q(1-H_i)$ is at least as big as the corresponding term $q(1-(H_i+w))$. This is because of the form of the function $q$. Note that $q$ is increasing in $[0,1]$ and also that the value of $q$ at any point $x\in [0,1]$ is greater than the value of $q$ at any point $y\in [-1,0)$. Therefore since $1-H_i>0$ and since we only subtract $w$ from each point, it follows that (\ref{Eq:w}) holds.

\noindent {\bf Case 2:} $\xi(\bu_4) = 1$

Assume without loss of generality that $\bu_4$ is $1$ on $P_1$ only. 
In this case, $H_1 = 0$, $H_2 = x+y$ and $H_3 = x+z$. The LHS of inequality (\ref{Eq:w}) is now: $LHS = q(1) + q(1-(x+y)) + q(1-(x+z))$, whereas the RHS is:
$$ RHS = q(1-w) + q(1-(x+y+w)) + q(1-(x+z+w)) = q(1-w) + q(-1+z)+ q(-1+y)$$ by using the fact that $x+y+w = 2-z$.

Let $\alpha_1 = 1$, $\alpha_2 = 1-(x+y)$, $\alpha_3 = 1-(x+z)$. The LHS is the sum of the values of $q$ at these points whereas the RHS is the sum of the values of $q$ after shifting each point $\alpha_i$ to the left by $w$. Let $\alpha'_i = \alpha_i-w$. The difference $\Delta = q(1) - q(1-w)$ will always be positive since $q(1)$ is the highest value that $q$ achieves in $[-1,1]$. Therefore to show that (\ref{Eq:w}) holds it is enough to show that the potential gain in $q$ from shifting $\alpha_2$ and $\alpha_3$ is at most $\Delta$. Suppose not and consider such a configuration. This means that either $q(\alpha'_2)>q(\alpha_2)$ or $q(\alpha'_3)>q(\alpha_3)$ or both. We will consider the case that both points achieve a higher value after being shifted. The same arguments apply if we have only one point that improves its value after subtracting $w$. Hence we assume that $q(\alpha'_2)>q(\alpha_2)$ and $q(\alpha'_3)>q(\alpha_3)$. Before we proceed, we state some properties of the function $q$, which can be verified by simple calculations:

\begin{claim}
\label{Cl:q}
The function $q$ is decreasing in $[-1,-\lambda]$ and increasing in $[-\lambda, 1]$.
Furthermore, for any $2$ points $x,y$ such that $x\in [-1,2-3\lambda]$ and $y\geq 2-3\lambda$, $q(y)\geq q(x)$.
\end{claim}

Using the above claim, we can argue about the location of $\alpha_2$ and $\alpha_3$. If $\alpha_2\geq 2-3\lambda\geq -\lambda$, then $q(\alpha_2)\geq q(\alpha'_2)$. Thus both $\alpha_2$ and $\alpha_3$ must belong to $[-1,2-3\lambda] = [-1, -1+\frac{3}{2t}]$. We will restrict further the location of $\alpha_2$ and $\alpha_3$ by making some more observations about $q$. The interval $[-1,2-3\lambda]$ is the union of $A_1 = [-1,-\lambda]$ and $A_2 = [-\lambda, 2-3\lambda]$ and we know $q$ is decreasing in $A_1$ and increasing in $A_2$. We claim that $\alpha_2, \alpha_3$ should belong to $A_1$ in the worst possible violation of (\ref{Eq:w}). To see this, suppose $\alpha_2\in A_2$ and $\alpha_3\in A_2$ (the case with $\alpha_2\in A_2$, $\alpha_3\in A_1$ can be handled similarly). We know that $q$ is the sum of a linear function and the function $x^{2t}$. Hence when we shift the $3$ points to the left, the difference $q(1) - q(1-w)$ is at least as big as a positive term that is linear in $w$. This difference has to be counterbalanced by the differences $q(\alpha'_2) - q(\alpha_2)$ and $q(\alpha'_3)-q(\alpha_3)$. However the form of $q$ ensures that there is a point $\zeta_2\in A_1$ such that $q(\alpha_2) = q(\zeta_2)$ and ditto for $\alpha_3$.
Hence by considering the configuration where $\alpha_2\equiv \zeta_2$ and $\alpha_3\equiv\zeta_3$ we will have the same contribution from the terms $q(\alpha'_2) - q(\alpha_2)$ and $q(\alpha'_3)-q(\alpha_3)$ and at the same time a smaller $w$. 

Therefore we may assume that $w\leq |A_1| = \frac{1}{2t}$, which is a very small number. 
By substituting the value of $q$, (\ref{Eq:w}) is equivalent to showing that:
$$ 1-(1-w)^{2t} + 6t\lambda^{2t-1}w \geq (\alpha_2-w)^{2t} - \alpha_2^{2t} + (\alpha_3-w)^{2t} - \alpha_3^{2t} $$

It is easy to see that the difference $1-(1-w)^{2t}$ is greater than or equal to the difference $(\alpha_2-w)^{2t} - \alpha_2^{2t}$. Hence it suffices to show:
$$ 6t\lambda^{2t-1}w \geq (\alpha_3-w)^{2t} - \alpha_3^{2t} $$
Since $w$ is small, we estimate the difference $(\alpha_3-w)^{2t} - \alpha_3^{2t}$ using the first derivative of $x^{2t}$ (the lower order terms are negligible). Thus the RHS of the above inequality is at most $2t|\alpha_3|^{2t-1}w$, which is at most $2tw$. But the LHS is:
$$ 6t\lambda^{2t-1}w \geq 6t/e w > 2tw$$
Therefore no configuration in this case can violate (\ref{Eq:w}) and we are done.

\noindent {\bf Case 3:} $\xi(\bu_4) = 2$

Assume that $\bu_4$ is $1$ on $P_1$ and $P_2$. 
Now $H_1 = y$, $H_2 = x$,
$H_3 = x+y+z$. The LHS and RHS of (\ref{Eq:w}) are now:
\begin{eqnarray*}
LHS &=& q(1-y) + q(1-x) + q(1-(x+y+z))\\
RHS & = & q(1-(y+w)) + q(1-(x+w)) + q(-1) 
\end{eqnarray*}

As in case 2, let $\alpha_1 = 1-y$, $\alpha_2 = 1-x$ and $\alpha_3 = 1-(x+y+z)$ be the $3$ points before shifting by $w$. First note that either $\alpha_1> 0$ or $\alpha_2> 0$. This comes from the constraint that $x+y+z+w = 2$. Assume that $\alpha_1> 0$. Hence $q(\alpha_1) - q(\alpha_1-w)>0$. If $\alpha_2\not\in[-1,2-3\lambda]$ then we would be done because by Claim~\ref{Cl:q}, $q(\alpha_2) - q(\alpha_2-w)>0$. Therefore the only way that (\ref{Eq:w}) can be violated is if the nonlinear term $(\alpha_3-w)^{2t} - \alpha_3^{2t}$ can compensate for the loss for the other terms. It can be easily checked that this cannot happen. Hence we may assume that both $\alpha_2,\alpha_3\in[-1,2-3\lambda]$ and that $q(\alpha_2-w)>q(\alpha_2)$, $q(\alpha_3-w)> q(\alpha_3)$. 
The rest of the analysis is based on arguments similar to case 2 and we omit it from this version. 

\noindent {\bf Case 4:} $\xi(\bu_4) = 3$
This case can also be done using similar arguments with case 2 and 3.
\end{proofof}

\subsection{A technical lemma}

\begin{lemma}
\label{calculus} The function
$f(x)=\frac{\alpha(x+1)+1/2}{1-2^{-x}}$ for $\alpha=\frac{\ln
2}{14 - 8\ln 2}$ attains its minimum in $[1,\infty]$ at $x=3$.
\end{lemma}
\begin{proof}
The derivative of $f$ is
$$\frac{1-2^{-x} - (\alpha(x+1)+1/2) \ln(2)
2^{-x}}{(1-2^{-x})^2}.$$
It is easy to see that $f'(3)=0$, $f(1)=4\alpha+1>8/7$, and
$\lim_{x \rightarrow \infty} f(x)=\infty$. So it is sufficient to
show that
$$g(x)=1-2^{-x} - (\alpha(x+1)+1/2) \ln(2)
2^{-x},$$
is an increasing function in the interval $[1,\infty)$. To show
this note that
$$g'(x)=2^{-x}\ln(2)\left(1-\alpha+\alpha
x\ln(2)+\alpha\ln(2)\right)>0,$$
for $x \ge 1$.
\end{proof}



\ignore{
\section{questions}
\paragraph{Low level: }
\begin{enumerate}
\item can we show that Theorem \ref{Thm:SDP-l1} holds in general
for relaxation that come from cuts and contain a norm constraint?

\end{enumerate}

\paragraph{High level: }
\begin{enumerate}
\item Are the m.s. above hypermetric?
\item can we get anything (like a constant above 1) against uniform sparsest
cut?
\end{enumerate}

\section{todo}
\begin{enumerate}
\item referring to Charikar's metric as $Y$?
\end{enumerate}
}

\end{document}